\documentclass[aps,pra,twocolumn]{revtex4}
\usepackage{graphicx}
\usepackage{bm}
\usepackage{amsmath,amssymb,amsthm,dsfont,bm}
\usepackage{color}
\usepackage{wasysym}

\usepackage[applemac]{inputenc}

\begin{document}

\title{Phase-space quantum Wiener-Khintchine theorem}
\author{Ainara \'Alvarez-Marcos and Alfredo Luis}
\email{alluis@fis.ucm.es}
\homepage{http://www.ucm.es/info/gioq}
\affiliation{Departamento de \'{O}ptica, Facultad de Ciencias
F\'{\i}sicas, Universidad Complutense, 28040 Madrid, Spain}
\date{\today}

\begin{abstract}
We derive a quantum version of the classical-optics Wiener-Khintchine theorem within the framework of detection of phase-space displacements with a suitably designed quantum ruler. A phase-pace based quantum mutual coherence function is introduced that includes the contribution of the detector. We obtain an universal equality linking resolution with coherence. This is illustrated with the case of Gaussian states and number states.     
\end{abstract}

\maketitle

Last years have witnessed the construction {\it ab initio} of purely quantum theories of coherence \cite{QC}. These approaches are based in abstract mathematical definitions that must fulfill some consistence  requirements under transformations. To some extent, these formulations contrast with the physical motivation and simplicity of the classical-optics coherence theory, as reflected for example in the Wiener-Khintchine theorem \cite{MW95}, being also divorced form the coherence theory of Glauber \cite{RJG63,PKJ19}. 

The aim of this work is to develop a contribution to the theory of coherence in quantum physics as a quantum translation of the  Wiener-Khintchine theorem. Since classical coherence is directly connected with interferometric resolution, we focus on the application of coherence to metrology showing that coherence is the actual resource behind quantum metrology. 

In Ref. \cite{AL21a} we have pursued the goal of translating the Wiener-Khintchine theorem to the quantum domain in the case of detection of signals encoded by uniparametric unitary transformations. In this work we extend this formalism to two-parameter unitary transformations in the form of displacements in phase space, which in our case is the complex-amplitude plane of a single-mode electromagnetic field. Displacements are basic operations in quantum optics, linked to the very same structure of the phase space.

The quantum version of the Wiener-Khintchine theorem presented here is built from first principles. The key ingredient is the introduction of a quantum ruler. This is a measuring scheme constructed as the collection of the effects caused by all possible displacements. The final form of the theorem emerges when using the Wigner-function formalism, because of its intimate connection with displacements and the structure of the phase space \cite{PWQa,PWQb}. The result has many desirable properties, including that the resolution does not depend on the signal value and an exact relation between resolution and quantum coherence.

\bigskip

Our system will be a single-mode electromagnetic field to be represented by the complex-amplitude operator $a$. The unitary displacement operator is 
\begin{equation}
    D(\beta) = e^{\beta a^\dagger - \beta^\ast a} ,
\end{equation}
as the transformation that imprints the signal to be detected, this is $\beta$, on the probe state $\rho_0$ as 
\begin{equation}
\label{rhobeta}
     \rho (\beta)  = D(\beta)\rho_0 D^\dagger (\beta) ,
\end{equation}
being $\beta$ any complex number.

\bigskip

The measurement to be performed to estimate the displacement $\beta$ will be represented by a positive operator-valued measure (POVM) $\Pi (\alpha)$, such that the conditional probability of obtaining the result $\alpha$ conditioned to a phase-space displacement $\beta$ is
\begin{equation}
\label{stat}
    p(\alpha|\beta) = \mathrm{tr} \left [ \rho (\beta ) \Pi (\alpha) \right ] .
\end{equation}
We will adapt $\Pi (\alpha)$ to the group of transformations we are studying inspired by the standard way of measuring lengths or times with rulers and clocks, for example. A ruler is constructed as the collection of the effects on a given reference element $\Pi_0$ caused by the action of the group of transformations we want to detect. This is 
\begin{equation}
\label{ds}
    \Pi (\alpha) = D (\alpha) \Pi_0 D^\dagger(\alpha), \quad \Pi_0 = \Pi (\alpha = 0) ,
\end{equation}
where the reference $\Pi_0 $ will be referred to as the tick or tick state. So, as reflected in Eq. (\ref{stat}), the detection is the comparison of the signal effect on the probe state with the collection of effects-marks that are stored in the ruler $\Pi (\alpha)$. {\it Grosso modo}, this ruler follows the idea that time is what you measure with a clock. A natural choice then will be that probe and tick state are proportional, but we will come to that later.  

\bigskip

The properties that a probability distribution $p(\alpha|\beta)$ must satisfy, these are reality, positivity, and normalization, translate into conditions on $\Pi(\alpha)$ in the form
\begin{equation}
\label{lpovm}
    \Pi (\alpha ) = \Pi^\dagger (\alpha ), \quad \Pi (\alpha ) \geq 0 , \quad \int d^2 \alpha \Pi( \alpha) = I ,
\end{equation}
where $I$ is the identity operator. After the ruler definition (\ref{ds}), these must imply some conditions on the tick $\Pi_0$. 

\noindent {\it Theorem.--} The conditions that $\Pi_0$ should satisfy are 
\begin{equation}
    \Pi_0^\dagger = \Pi_0, \quad \Pi_0 \geq 0, \quad  \mathrm{tr} \Pi_0 = \frac{1}{\pi}. 
\end{equation}
This is to say that, apart from the normalization factor $\pi$, we get that the tick must be a density matrix, without any further restriction.

\noindent {\it Proof.--} The first two requirements are immediate after Eqs. (\ref{ds}) and (\ref{lpovm}). For the last one we construct the $\mathcal{I}$ operator
\begin{equation}
\mathcal{I} = \int d^2 \alpha  D (\alpha) \Pi_0 D^\dagger(\alpha) , 
\end{equation}
and we check whether $\mathcal{I} \propto I$. To this end we use the continuous basis $|x\rangle$ of eigenstates of the quadrature $X$, where $X =( a + a^\dagger)/2$, $Y= i ( a^\dagger - a  )/2$, are the quadrature operators, that satisfy the typical commutation law of position and linear momentum $[X,Y] = i/2$. Furthermore we will use that 
\begin{equation}
    D(\alpha) = e^{2i(\alpha_y X-\alpha_x Y)} = e^{-i\alpha_y \alpha_x} e^{2i\alpha_y X}e^{-2i\alpha_x Y} ,
\end{equation}
and 
\begin{equation}
    e^{2i\alpha_x Y} |x \rangle =  |x - \alpha_x \rangle  ,
\end{equation}
where $\alpha_{x,y}$ are the real and imaginary parts of $\alpha$ as $\alpha= \alpha_x+i\alpha_y$. With all this we construct the matrix elements of $\mathcal{I}$ in the $|x \rangle$ basis to get
\begin{equation}
    \langle x | \mathcal{I} |x^\prime \rangle = \int d\alpha_y e^{2i \alpha_y (x-x^\prime )} \int d\alpha_x  \langle x -\alpha_x| \Pi_0 |x^\prime - \alpha_x\rangle,
\end{equation}
leading to $\langle x | \mathcal{I} |x^\prime \rangle = \pi \delta (x-x^\prime ) \mathrm{tr} \Pi_0 $, and then finally, the last relation in Eq. (\ref{lpovm}) implies that $\mathrm{tr} \Pi_0 = 1/\pi$. $\blacksquare$

\bigskip

Once the measurement has been planned and carried out, metrology deals with extracting certainties about the unknown $\beta$ from the conditional statistics $ p(\alpha|\beta)$. After Eqs. (\ref{rhobeta}), (\ref{stat}) and (\ref{ds}) we get that the functional form of the conditional statistics does not depend on $\beta$, we may say it is shift-invariant
\begin{equation}
\label{yes}
     p(\alpha|\beta) = p(\alpha -\beta) , \quad
     p(\mu) = \mathrm{tr} \left [ \rho_0  D(\mu) \Pi_0 D^\dagger (\mu) \right ] ,
\end{equation}
meaning that the properties of the setup are independent of the unknown $\beta$. More specifically, since $p(\mu)$ depends just on $\rho_0$ and $\Pi_0$ that are assumed to be known before the measurement, we have that the resolution in the detection process can be known in advance, being the same for all signals $\beta$, as a good property of the so constructed ruler. 

\bigskip

We will find convenient to express the main result in terms of the Wigner-Weyl correspondence between Hilbert-state operators and functions on the phase space, in our case the complex plane \cite{PWQa,PWQb}. The correspondence is realized after a double relation between any operators $A$ and functions $W_A (\alpha)$ 
\begin{equation}
\label{WO}
W_A (\alpha) = \mathrm{tr} \left [ A \Omega (\alpha) \right ], \quad A = \pi \int d^2 \alpha W_A (\alpha ) \Omega (\alpha) ,
\end{equation}
being
\begin{equation}
\Omega (\alpha) = \frac{1}{\pi^2} \int d^2 \eta \, e^{\alpha \eta^\ast - \alpha^\ast \eta} D (\eta ) , 
\end{equation}
or equivalently
\begin{equation}
\label{DO}
D(\beta) = \int d^2 \alpha e^{\alpha^\ast \beta - \alpha \beta^\ast} \Omega (\alpha) .
\end{equation}
Among its properties we have $W_{A^\dagger} (\alpha)=W^\ast_A (\alpha)$ after $\Omega^\dagger (\alpha) = \Omega (\alpha)$, the possibility of computing expected values as integrals over the phase space:
\begin{equation}
\label{trsp}
\mathrm{tr} (AB) = \pi \int d^2 \alpha W_A (\alpha ) W_B (\alpha ) ,
\end{equation}
and the transformation law under displacements
\begin{equation}
\label{td}
\rho (\beta) = D(\beta) \rho_0 D^\dagger (\beta) \rightarrow W_\beta (\alpha) = W_0 (\alpha - \beta ),
\end{equation}
exactly like a classical distribution over phase space, that is, with a transformation of its arguments.

\bigskip

With all this we can finally demonstrate the main result presented in this work. To this end we assume that both the probe $\rho_0$ and the POVM tick $\Pi_0$ are pure states,
\begin{equation}
\label{ps}
\rho_0 = |\psi \rangle \langle \psi |, \quad \Pi_0 = \frac{1}{\pi}|\phi \rangle \langle \phi | .
\end{equation}
Mixed states only add classical subjective randomness that deep down may be treated with the tools of classical optics. Under these assumptions, the statistics is given by \begin{equation}
\label{pmsdo}
    p(\mu) = \frac{1}{\pi} \left | \langle \psi | D(\mu ) | \phi \rangle \right |^2 ,
\end{equation}
that lead us to the main result of this work. 

\bigskip

\noindent {\it Theorem.--} The  statistics $p(\mu)$ in Eqs. (\ref{yes}) and (\ref{pmsdo}) obtained via a quantum ruler with probe and tick being pure states satisfies a phase-space version of the Wiener-Khintchine theorem as a symplectyc Fourier transform of a mutual coherence function $\Gamma (\tau)$:
\begin{equation}
\label{QWKT}
    p(\mu) = \int d^2 \tau e^{\tau \mu^\ast - \tau^\ast \mu} \Gamma (\tau ) ,
\end{equation}
where 
\begin{equation}
\label{CF}
\Gamma (\tau) = \frac{1}{\pi} \int d^2 \alpha W^\ast_{|\psi \rangle \langle \phi |} (\alpha) W_{|\psi \rangle \langle \phi |} (\alpha + \tau) ,
\end{equation}
where $W_A (\alpha)$ is the corresponding Wigner function, or equivalently 
\begin{equation}
\label{ChF}
\Gamma (\tau) =  \frac{1}{\pi^2} \mathcal{C}^\ast_\psi  (\tau) \mathcal{C}_\phi (\tau) ,
\quad  \mathcal{C}_\psi  (\tau) = \langle \psi | D (\tau ) |\psi \rangle, 
\end{equation}
where $\mathcal{C}_\psi  (\tau)$ is the symmetrically ordered characteristic function of the corresponding state, and equivalently for $\mathcal{C}_\phi$ \cite{PWQa,PWQb}. 

\bigskip 

\noindent {\it Proof.--} We start form the second relation in Eq. (\ref{yes}) with the choice of probe and tick in Eq.  (\ref{ps}). Then, using twice Eq. (\ref{DO}) for $D(\mu)$ followed by the application twice of the first relation in Eq. (\ref{WO}) applied to $A=|\psi \rangle \langle \phi |$ and taking into account that $W_{A^\dagger} (\alpha ) = W^\ast_{A} (\alpha )$ leads to Eqs. (\ref{QWKT}) and (\ref{CF}) after a simple change of variables defining $\tau$. Finally, after Eq. (\ref{CF}) and using Eqs. (\ref{WO}) and (\ref{td}) we get 
\begin{equation}
\Gamma (\tau) = \frac{1}{\pi} \int d^2 \alpha W_{|\phi \rangle \langle \psi |} (\alpha) W_{D^\dagger (\tau)|\psi \rangle \langle \phi |D (\tau)} (\alpha ) ,
\end{equation}
so that after Eq. (\ref{trsp}) we readily get (\ref{ChF}). $\blacksquare$ 

\bigskip

We see that the mutual coherence function $\Gamma (\tau)$ emerges as an auto-correlation function of the the Wigner function of the combined probe-tick state, much in the same way that the optical coherence function in classical optics emerges in terms of the auto-correlation of the electric field, for example when expressed as a time average. On the other hand, characteristic functions are completely measurable quantities. Quantum tomography is based on the measurement of $\mathcal{C}_\psi  (\tau)$ \cite{VR89}, and there are also other proposals of directly addressing the measurement of characteristic functions as shown in Refs. \cite{FH20,ZPM04} for example. Finally, we think it is worth noting the explicit appearance of the apparatus state which is a good property of this formulation in agreement with the peculiarities of quantum observation and measurement. Among the properties of $\Gamma (\tau)$  it can be easily seen that $\Gamma^\ast (\tau) = \Gamma (- \tau)$ and $\Gamma (0) = 1/\pi^2$.

\bigskip

In the particular and natural case that the probe equals the tick, this is $|\phi \rangle = |\psi \rangle$ in Eq. (\ref{ps}), we have a much more simple expression 
\begin{equation}
\label{acf}
\Gamma (\tau) = \frac{1}{\pi} \int d^2 \alpha W_\psi (\alpha) W_\psi (\alpha + \tau) = \frac{1}{\pi^2} \left | \mathcal{C}_\psi  (\tau) \right |^2 ,
\end{equation}
and a maybe unexpected result.

\bigskip

\noindent {\it Theorem.--} If the probe equals the tick being pure states, i. e., $|\phi \rangle = |\psi \rangle$ in Eq. (\ref{ps}), then statistics and coherence function are proportional $p(\mu) = \pi \Gamma (\mu )$.

\bigskip 

\noindent {\it Proof.--} Using Eqs. (\ref{trsp}) and (\ref{td}) in Eq. (\ref{pmsdo}) readily leads to Eq. (\ref{acf}) for $p(\mu) / \pi$ instead of $\Gamma (\mu )$ and $\tau = \mu$. $\blacksquare$

\bigskip

So we have the curious unexpected result that coherence and the statistics are identical being eigenfunctions of the symplectyc Fourier transform. To understand this we may consider that $p(\mu)$ is not a typical probability distribution given by projection on orthogonal states, where coherences are removed reducing the density matrix to its diagonal in the base of the measured observable. By construction, $\Pi (\alpha)$ is not the projection onto orthogonal states so it preserves the coherences. This is clearly illustrated by the first example to be considered next, where $p(\mu)$ is the $Q$ function of $\rho_0$, so it contains complete information about $\rho_0$ including its coherence properties.

\bigskip

Seemingly this result dose not extend to other orderings of the $a$, $a^\dagger$ operators, different from the symmetric case. This is because the corresponding normal and anti-normal characteristic functions are related to the symmetrical ordering by multiplication by a real exponential, while the statistics does not depend on the operator ordering. A different case arise regarding ordering with respect $X$ and $Y$ quadratures instead of $a$, $a^\dagger$, since in such a case the modulus of the characteristic functions are the same for the orderings  $X-Y$, $Y-X$ and symmetric, but this kind of orderings are rarely considered.

\bigskip

The characteristic-function form for $\Gamma (\tau) $ in Eqs. (\ref{ChF}) and (\ref{acf}) is particularly simple. Characteristic functions carry complete information about the corresponding state or observable, in much the same way that the Wigner-Weyl representation does. So it is not surprising that they can determine the statistics $p(\mu)$ is some way or the other. The opposite direction might be more surprising, this is that the statistics $p(\mu)$ may serve to infer, at least partially, the characteristic function of the probe and tick states. 

\bigskip

By linearity, the case in which the probe or the tick are not pure leads to a $\Gamma (\tau)$ function of the form 
\begin{equation}
\label{mix1}
    \Gamma (\tau) = \sum_{j,k} p_j p^\prime_k \Gamma_{j,k} (\tau ) ,
\end{equation}
where 
\begin{equation}
\label{mix2}
\Gamma_{j,k} (\tau ) = \frac{1}{\pi} \int d^2 \alpha W_{|\phi_k \rangle \langle \psi_j |} (\alpha) W_{|\psi_j \rangle \langle \phi_k |} (\alpha +\tau ) ,
\end{equation}
being $|\psi_j \rangle$ the eigenvector of $\rho_0$ with eigenvalue $p_j$ and  $|\phi_k \rangle$ the eigenvector of $\Pi_0$ with eigenvalue $p^\prime_k$, and the sum extends to all eigenvectors. As a further consequence we get that the case $\rho_0 \propto \Pi_0$ is not given by an straightforward generalization of Eq. (\ref{acf}). Note that in such a case linearity is lost. 

\bigskip

Finally we can derive a natural resolution-coherence relation implied by this theory. In general terms, being $p(\mu)$ and $\Gamma (\tau )$ a Fourier transform pair, the Parseval's theorem applies and we may establish the following coherence-resolution relation 
\begin{equation}
\label{tcdb}
\tau_c  \Delta \beta = \frac{1}{\pi^2} ,
\end{equation}
where $\tau_c$ represents the phase-space analog of the  coherence time, and  $\Delta \beta$ a R\'enyi-type measure of uncertainty regarding the variable $\mu$ obeying statistics $p(\mu)$ 
\begin{equation}
\label{inc}
    \tau_c = \int d^2 \tau |\Gamma (\tau ) |^2 , \quad \Delta \beta = \frac{1}{ \int d^2 \mu \, p^2 (\mu)} .
\end{equation}
We may consider $\Delta \beta$ a measure of uncertainty or resolution in the determination of the signal $\beta$, so that Eq. (\ref{tcdb}) is actually a connection between resolution and coherence. 

\bigskip

These measures of coherence and resolution may seem somewhat exotic, but they have been already presented and analyzed in the domain of classical-optics coherence \cite{MW95}, as well as in the quantum domain as useful measures of quantum uncertainty \cite{CBTW17,GZ01}.

We may say that the meaning of {\it coherence time} in this phase-space plane context is fully equivalent to its very intuitive meaning in classical optics, so that larger coherence means larger resolution. This is to say that coherence acts as a resource for quantum resolution in the same terms it is a resource for classical-optics interferometry.    

\bigskip

We can develop further the connection of $\Gamma (\tau)$ with some other current techniques in quantum metrology in the case $|\psi \rangle = | \phi \rangle$. To this end let us consider that the displacement occurs along an specific direction in the complex $\tau$ plane determined by the unit real vector $\boldsymbol{n}$. Let us introduce the following measure of coherence
\begin{equation}
\label{cm}
  T_c (\boldsymbol{n}) =  \int d^2 \tau (\boldsymbol{\tau}\cdot \boldsymbol{n} )^2 \Gamma (\tau )  = 2 \boldsymbol{n}^T \boldsymbol{C} \boldsymbol{n} ,
\end{equation}
where $\boldsymbol{\tau}= (\tau_x,\tau_y)$, the subscript $T$ denotes transposition, and $\boldsymbol{C}$ is the $2 \times 2$ quadrature symmetrically ordered covariance matrix, this is $C_{x,x}= \Delta^2 X$, $C_{y,y} = \Delta^2 Y $, and
\begin{equation}
\label{cm2}
C_{x,y} =C_{y,x} = \frac{1}{2}\langle \psi | \left ( XY+YX \right ) |\psi \rangle - \langle \psi | X |\psi \rangle\langle \psi | Y |\psi \rangle .
\end{equation}
It is known that the inverse of the covariance matrix is a lower bound to the corresponding Fisher matrix, being an equality for Gaussian statistics \cite{KL99}.

\bigskip

To demonstrate this result we begin with 
\begin{equation}
    T_c (\boldsymbol{n}) = 2 \sum_{j,k=x,y} n_j n_k C_{j,k} , \quad 
    C_{j,k} = \frac{1}{2}  \int d^2 \tau \, \tau_j \tau_k \Gamma (\tau ) ,
\end{equation}
using the Wigner-function for $\Gamma (\tau )$ in Eq. (\ref{acf}). After a suitable change of variables 
\begin{equation}
       C_{j,k} = \frac{1}{2\pi}  \int d^2 \alpha d^2 \alpha^\prime \, \left  ( \alpha^\prime_j - \alpha_j \right )\left  ( \alpha^\prime_k - \alpha_k \right ) W_\psi (\alpha ) W_\psi (\alpha^\prime ) .
\end{equation}
The final result follows by noting that the Wigner function of the powers of the quadrature operators are simply, 
\begin{equation}
    W_{X^n} (\alpha) = \alpha_x^n ,\quad   W_{Y^n} (\alpha) = \alpha_y^n , 
\end{equation}
because the  marginals of $W_\psi (\alpha)$ are the exact probability distributions for the corresponding quadrature, and  
\begin{equation}
    W_{\frac{1}{2} (XY+YX)} (\alpha) =  \alpha_x \alpha_y ,
\end{equation}
in accordance to symmetric ordering. Then, Eqs. (\ref{cm}) and (\ref{cm2}) hold  after the proper application of the property in Eq. (\ref{trsp}). The case in which $|\psi \rangle \neq | \phi \rangle$ is not so transparent and the elements of the matrix $C$ are of the form
\begin{equation}
    C_{x,x}= \operatorname{Re} \left \{ \langle \phi |X^2 |\psi \rangle \right \} - \left |\langle \phi |X |\psi \rangle \right |^2 ,
\end{equation}
\begin{equation}
    C_{y,y}= \operatorname{Re} \left \{ \langle \phi |Y^2 |\psi \rangle \right \} - \left |\langle \phi |Y |\psi \rangle \right |^2 ,
\end{equation}
and 
\begin{equation}
C_{x,y} = \operatorname{Re} \left \{ \frac{1}{2}\langle \phi | \left ( XY+YX \right ) |\psi \rangle - \langle \phi | X |\psi \rangle\langle \psi | Y |\phi \rangle \right \}.
\end{equation}
Therefore, after Eqs. (\ref{mix1}) and (\ref{mix2}) there seems to be no straightforward generalization to mixed states.

\bigskip

Regarding the relation with previous results, we can show that the mutual coherence function in Eq. (8) of \cite{AL21a} when the generator is a quadrature, say $G=X$, referred to here as  $\Gamma_G (\tau)$ to avoid confusions, directly emerges as the corresponding marginal of the mutual coherence function $\Gamma (\tau_x,\tau_y)$ introduced in this work in Eq.  (\ref{CF}), this is \begin{equation}
    \Gamma_G (\tau) = \int d \tau_y  \Gamma (\tau_x=\tau,\tau_y) .
\end{equation}
This holds because the marginal of the Wigner function provides the exact statistics for the corresponding quadrature. The other way round, this  suggests that the equivalence between the two approaches may be implemented for arbitrary groups of transformations by a suitable generalization of Wigner-Weyl correspondence, such as the SU(2) Wigner function for example \cite{VG89}.

\bigskip

We can illustrate these results with two simple but meaningful examples. To begin with we consider $|\psi \rangle = |\phi \rangle$ to be the squeezed vacuum, whose Wigner function is
\begin{equation}
     W(\alpha ) = \frac{2}{\pi} e^{- 2 x^2/\lambda - 2 \lambda y^2} , \quad \alpha = x + i y,
\end{equation}
where  $\lambda$ is expressing quadrature squeezing  
\begin{equation}
\Delta^2 X = \frac{\lambda}{4}, \quad \Delta^2 Y = \frac{1}{4 \lambda},
\end{equation}
being for all $\lambda$ minimum-uncertainty states $\Delta X \Delta Y = 1/4$. In this case we get
\begin{equation}
     \Gamma(\tau ) = \frac{1}{\pi^2} e^{- \tau_x^2/\lambda - \lambda \tau_y^2} , \quad \tau = \tau_x + i \tau_y .
\end{equation}
After Eq. (\ref{pmsdo}) it is easily seen that $p(\mu)$ is a squeezed version of the Husimi $Q$ function, being exactly the $Q$ function when $\lambda = 1$. In experimental terms, this corresponds to the statistics of the double homodyne detection, with unbalanced input beam splitter for $\lambda \neq 1$. The measure of coherence defined in Eq. (\ref{inc}) gives $\tau_c =1/(2 \pi^3)$, that does not depend at all on $\lambda$.

\bigskip

Let us consider an example where $|\psi \rangle \neq |\phi \rangle$, being both squeezed vacuum states with different squeezing parameters, say $\lambda$ and $\mu$, respectively. In such a case we readily obtain
\begin{equation}
     \Gamma(\tau ) = \frac{1}{\pi^2} e^{- (\lambda + \mu)\tau_x^2/(2 \lambda \mu) - (\lambda + \mu)\tau_y^2/2} , 
\end{equation}
showing explicitly the symmetric probe and detector contributions, leading to a coherence time 
\begin{equation}
    \tau_c = \frac{1}{\pi^3} \frac{\sqrt{\lambda/\mu}}{1 +\lambda/\mu} ,
\end{equation}
that reaches its maximum when $\lambda = \mu$.

\bigskip

\begin{figure}[h]
    \includegraphics[width=7cm]{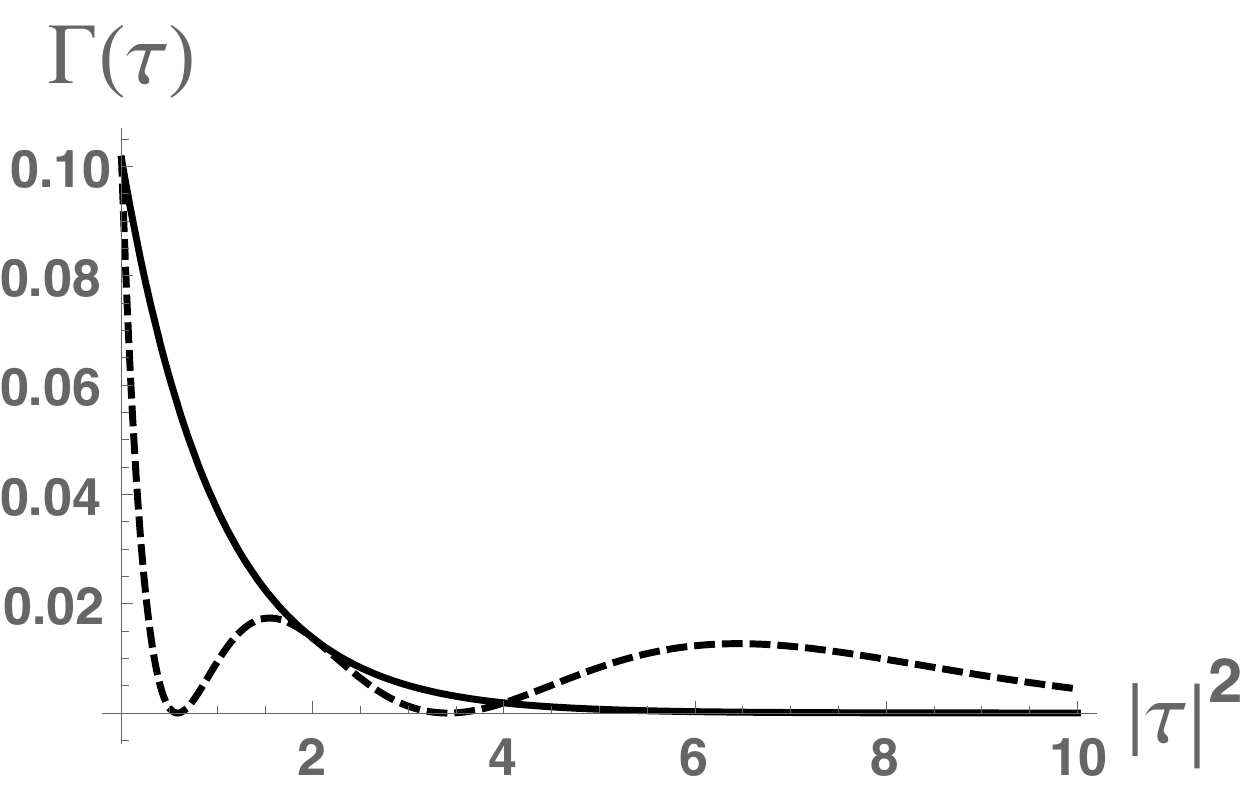}
    \caption{Plots of $\Gamma (\tau)$ for $n=0$ (solid line) and $n=2$ (dashed line), as functions of $|\tau|^2$.}
    \label{Gamma}
\end{figure}

Finally, let us now consider that probe and tick are photon-number states with $n$ photons, that is $|\psi \rangle = |\phi \rangle = | n \rangle$, that in the vacuum case $n=0$ coincides with the $\lambda = 1$ Gaussian example above. In this case the Wigner function and mutual coherence functions read \cite{OKKB90}
\begin{equation}
     W(\alpha ) = \frac{2(-1)^n}{\pi} e^{-2 |\alpha|^2} \mathcal{L}_n (4 |\alpha |^2 ) ,
\end{equation}
and
\begin{equation}
    \Gamma (\tau) = \frac{e^{- |\tau|^2} }{\pi^2} \left [ \mathcal{L}_n (|\tau |^2 ) \right ]^2 ,
\end{equation}
respectively, where $\mathcal{L}_n$ are the Laguerre polynomials. In Fig. \ref{Gamma} we plot two examples of $\Gamma (\tau)$ for $n=0$ (solid line), which is also the Gaussian example above, and $n=2$ (dashed line), as functions of $|\tau|^2$. 

\begin{figure}[h]
    \includegraphics[width=7cm]{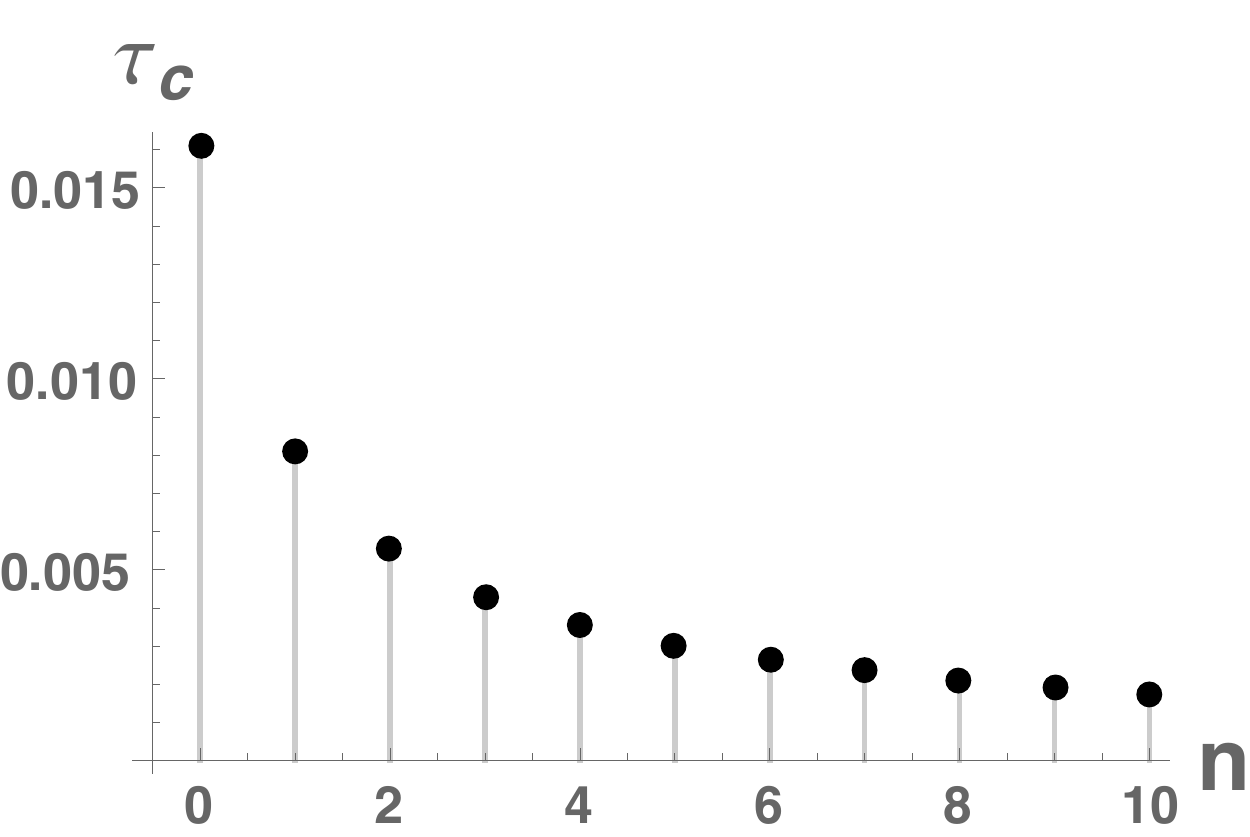}
    \caption{Plot of the coherence estimator $\tau_c$ as a function of the number of photons $n$ of the probe and tick. }
    \label{tc}
\end{figure}

We can appreciate in Fig. \ref{Gamma} that for the number case $\Gamma (\tau)$ has the strong oscillatory behavior typical of the number states, which is present both in the quadrature wave-function as well as in its Wigner function. In the long run, these oscillations imply a larger spreading that spoils coherence, as shown in the decreasing behavior of the coherence time $\tau_c$ for increasing $n$ in Fig. \ref{tc}. A similar result is clearly displayed by $T_c$ in Eq. (\ref{cm}) for any direction $\boldsymbol{n}$, since for number states all quadrature variances increase with increasing $n$. 

\bigskip

{\it Acknowledgments:} We thank both reviewers of the Optics Letters journal for their expert and thorough analysis of this proposal and very valuable suggestions that certainly improved the quality of the paper. This arXiv version includes some of the last recommendations that were not included in the journal version.

\end{document}